# Quantum states and intertwining phases in kagome materials

Yaojia Wang[1,2,4*], Heng Wu[1,2,4], Gregory T. McCandless[3], Julia Y. Chan[3], Mazhar N. Ali[1,2*]

[1]Kavli Institute of Nanoscience, Delft University of Technology, Lorentzweg 1, 2628 CJ , Delft, The Netherlands
[2]Department of Quantum Nanoscience, Faculty of Applied Sciences, Delft University of Technology, Lorentzweg 1, 2628 CJ , Delft, The Netherlands
[3]Department of Chemistry and Biochemistry, Baylor University, Waco, TX, 76798, USA
[4]These authors contributed equally

*Corresponding author: whyjwang@gmail.com, maz@berkeley.edu

**Abstract**: In solid materials, nontrivial topological states, electron correlations, and magnetism are central ingredients for realizing quantum properties, including unconventional superconductivity, charge and spin density waves, and quantum spin liquids. The Kagome lattice, made up of connected triangles and hexagons, can host these three ingredients simultaneously and has proven to be a fertile platform for studying diverse quantum phenomena including those stemming from the interplay of these ingredients. In this review, we introduce the fundamental properties of the Kagome lattice as well as discuss the complex observed phenomena seen in several emergent material systems such as the intertwining of charge order and superconductivity in some Kagome metals, modulation of magnetism and topology in some Kagome magnets, and symmetry breaking with Mott physics in the "breathing" Kagome insulators. We also highlight many open questions in the field as well as future research directions of Kagome systems.

In solid state systems, basic electronic states like metallic and semiconducting/insulating behavior have been well described by free or nearly-free electron theory based on non-interacting electrons in a periodic lattice potential. However for correlated systems, where the Coulomb interaction cannot be ignored and is comparable or stronger than the kinetic energy of the electrons, the interaction between particles (i.e. electron-electron and electron-phonon interactions) plays an important role in the material's physical properties. Electron correlation can induce instabilities at the Fermi surface and drive electronic fluctuations which can lead to many novel quantum phases with broken symmetry (Box 1), including superconductivity (SC), complex magnetism, charge density waves (CDW), spin density waves (SDW), nematic/smectic orders, pair density waves (PDW)[1–4], Mott insulating behavior, and more. These various phases have been observed in strongly correlated systems such as heavy fermion materials, high temperature superconductors, complex oxide perovskites, transition metal dichalcogenides, and twisted graphene[3,5–8]. However they need not be distinct; they can compete and even become entangled with each other in the same material, such as in the cuprate superconductors where superconductivity competes with CDW and SDW phases, and an observed PDW phase is proposed to be the possible "mother phase"[1]. The intertwining and interdependency of the different phases is complex and not yet fully understood, but may yield insight into understanding exotic physical properties like unconventional superconductivity.

The kagome lattice, a.k.a. trihexagonal tiling, is composed of hexagons and triangles in a network of corner-sharing triangles, and is an ideal structure for studying the intertwining of electron correlation and various quantum phases. It naturally hosts geometrical frustration due to its triangular lattice which, for example, causes magnetic frustration in some kagome magnets (Herbrtsmithite)[9,10] and has been predicted to give rise to a valence bond solid state[11] as well as a strongly correlated quantum spin liquid state[12]. Also the electronic interference between the three sublattices in the kagome net yields many electronic band structure features including topological Dirac/Weyl/nodal line points, van Hove singularities (vHSs), and flat bands. The vHSs and flat bands possess high densities of states and enhanced effective masses, resulting in strong correlations near these bands. Hence kagome materials are particularly ideal for studying the interplay of topology, magnetism and electron correlation.

Early studies of kagome metals have shown several properties, such as Weyl fermions in the ferromagnet $Co_3Sn_2S_2$[13,14], Dirac points and flat bands in $FeSn$[15] (antiferromagnet) and $CoSn$ (paramagnet)[16,17], signs of magnetic skyrmions in the noncollinear ferromagnet $Fe_3Sn_2$[18,19], chiral spin structures in the non-collinear antiferromagnet $Mn_3X$ (X=Sn, Ge)[20,21], and large anomalous Hall effects in many of these magnets[22,23]. Recently, several new material systems have been found which showcase both distinct quantum phases and topological states as well as the entanglement of these different states. An example system are the topological kagome metals $AV_3Sb_5$ (A=K, Cs, Rb)[24–27], which have attracted extreme research attention due to the observation of superconductivity, CDW, nematic/stripe order, PDW phase[28–31], and large anomalous Hall effect[32–34] all in a single material as a function of temperature. These complex symmetry breaking ordered states are found to compete or intertwine with each other, much like the situation of the high temperature superconductors. And now beyond $AV_3Sb_5$, new physical properties and possibilities are being explored in novel kagome systems: the $AM_6X_6$ (e.g. $TbMn_6Sn_6$)[35,36] family (a.k.a. the "166" compounds) has large chemical tunability and shows an array of magnetic phases (unlike the $AV_3Sb_5$ family which doesn't exhibit long-range magnetic order), while another family, $Nb_3X_8$ (X = Cl, Br, I) has trigonally distorted (a.k.a. has a breathing mode) kagome lattice which have prominent isolated flat bands and are proposed to be possible Mott insulators or obstructed atomic insulators[37–40].

In this review, we discuss the electronic and magnetic properties of kagome lattices, particularly highlighting several new material systems with diverse of features, including the $AV_3Sb_5$ family, "166" compounds and "0.5-3-3" (i.e. the hybrid $Yb_{0.5}Co_3Ge_3$)[41,42], and the recent distorted kagome materials ($Nb_3X_8$). The intertwining of charge orders and the superconducting state, with related studies of physical mechanisms behind these phases, will be discussed in $AV_3Sb_5$. The diverse magnetic phases and the modulation of magnetism under the influence of both inter-layer and intra-layer magnetic coupling are discussed in $AM_6X_6$, as well as the new topological states based on the interaction of magnetism and topology as well as the structural transition with possible CDW states in materials with weak magnetism. Distinct from the topological kagome metals, new properties in the breathing kagome system are discussed in $Nb_3X_8$ including a Mott insulating state with an isolated flat band, as well as a theorized obstructed atomic insulating state with layer dependent magnetism. Finally, we discuss future directions of investigation based on further kagome materials.

---
**BOX 1: Electron correlation related quantum phases**
- *Charge density wave (CDW) order*: a charge order that breaks translational symmetry. It normally causes a periodic lattice distortion due to electron-phonon coupling in systems with Fermi surface nesting (large, parallel sections of a Fermi surface that share a single vector in momentum space) and results in a splitting of energy states (i.e. a gap opening at the Fermi level) that lowers the ground state energy[2,43]. Unconventional CDW mechanisms are also proposed in some 2D and 3D materials (e.g. $TiSe_2$), driven by an exitonic insulator mechanism or a band Jahn-Teller effect[2].
- *Spin density wave (SDW) order*: These share many similarities with CDW order, including the breaking of translational symmetry and Fermi surface nesting, but arise from a periodic modulation of the spin density (rather than the charge density) and occur due to electron-electron interactions[44].
- *Nematic/smectic phases*: A nematic phase spontaneously breaks only rotational symmetry and tends **to have orientational ordering;** a smectic phase is a unidirectional stripe-like phase that breaks rotational symmetry as well as translational symmetry along one direction[45,46].
- *Pair density wave (PDW) phase:* a special superconducting state, where the amplitude of the superconducting pairing oscillates periodically in space so that the spatial average of the superconducting order vanishes[47]. It is expected to arise where superconductivity coexists with CDW or SDW order.

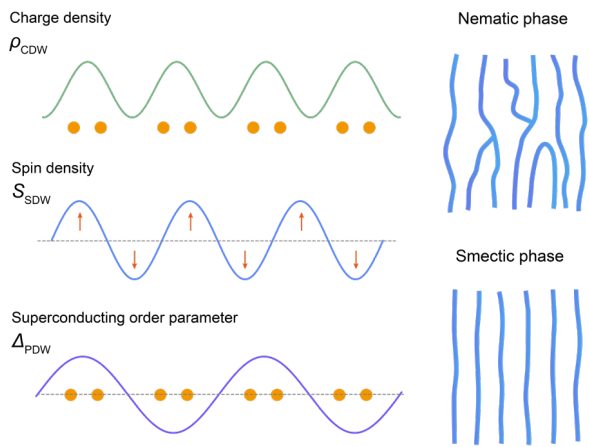

**Figure 1**. Schematic of different phases. The CDW, SDW and PDW phase corresponds to the periodic oscillation of charge density, spin density, and superconducting order parameter, respectively. The charge density in nematic and smectic phases tend to be unidirectional, and the schematics are drawn based on electronic liquid-crystal phases.

**The Kagome lattice and key electronic band structure properties**

The ideal kagome lattice is made up of corner-sharing triangles (Figure 2a), in which there are three lattice sites per unit cell (labeled as A, B, C sublattices). These sublattices induce a high degree of geometrical frustration from balanced exchange for electron hopping between the different atoms. The electron interaction in the kagome lattice can be described with a Hubbard model and the typical Hamiltonian takes the form[48,49]

$$H = H_0 + H_{\text{int}}$$

$$H_0 = \sum_{\{i,j\}\sigma} \left(t_{ij}\hat{c}_{i\sigma}^+ \hat{c}_{j\sigma} + h.c.\right) + \mu \sum_{i,\sigma} n_{i,\sigma}$$

$$H_{\text{int}} = U_0 \sum_i n_{i\uparrow} n_{i\downarrow} + \frac{U_1}{2} \sum_{\langle i,j \rangle, \sigma, \sigma'} n_{i,\sigma} n_{j,\sigma'}$$

$H_0$ is the Hamiltonian part of tight-binding model including electron hopping and the chemical potential ($\mu$). $t_{ij}$ is the hopping integral and represents the kinetic energy of electrons hopping between nearest-neighbor sites. $\hat{c}_{i\sigma}$ is the annihilation operator of electrons on the $i$ site with spin $\sigma$, and $n_{i,\sigma} = \hat{c}_{i\sigma}^+ \hat{c}_{i\sigma}$. $H_{\text{int}}$ is the Hamiltonian considering the on-site (local) Hubbard interaction of scale $U_0$ (local interaction), and nearest neighbor Coulomb interaction of scale $U_1$. Longer range interactions (e.g. next nearest neighbor interactions) can also be included in some systems. In the following, we discuss the general properties of "kagome bands", the three bands which are popularly associated with kagome lattice (Figure 2) and the influence of other factors on those bands and their corresponding physical properties.

Without including electron interactions and symmetry breaking, the basic band structure of the ideal kagome lattice ($t_{ij}$ is the same for nearest neighbor sites in upward and downward triangles, e.g. $t_1 = t_2$ in Figure 2a) based on a tight bonding model shows the typical features of: a doubly degenerated flat band connected with other bands, a Dirac point around the K point, and van-Hove singularities around the M point (Figure. 2c)[48–50]. The density of states (DOS) and electron effective mass at the Fermi level are highly elevated near the flat band and vHSs, where the strong correlation of the electrons needs to be considered (Figure. 2c). In addition, the hexagonal Fermi surface in momentum space is favorable to Fermi surface nesting where a vector can connect parallel Fermi surface sheets with three nesting vectors (Figure. 2b). This Fermi surface nesting is strong near the vHSs point, which creates a diverging electronic susceptibility and Fermi surface instability.

Upon breaking of symmetry, these kagome bands can be modulated and can generate new nontrivial topological states. One typical symmetry breaking kagome lattice distortion is the breathing kagome lattice[51,52] (Figure 2d); the breaking of inversion symmetry based on lattice trimerization gives rise to different hopping integrals between the 3 sites (the hopping amplitude is different for nearest neighbor sites in upward and downward triangles with $t_1 \neq t_2$ in Figure 2d), which opens a gap near the Dirac point in the band structure (Figure 2e). In the ideal kagome lattice, introducing strong Rashba spin-orbital coupling (SOC) or intrinsic SOC can also open a gap around the Dirac point and induce an isolated flat band Figure 2e[51]. On the other hand, time reversal breaking (e.g. ferromagnetic order) associated with SOC in the kagome lattice can also generate the isolated flat band with a large gap and nonzero Chern number, and a topologically nontrivial phase based on gap opening of the Dirac/Weyl point[53]. Isolated flat bands are of critical importance for unconventional quantum states; for instance, partially filling (such as 1/3 or 1/2 filled) of a flat band with non-zero Chern number is promising to realize a high temperature fractional quantum Hall effect[54]. The flat band with strong electron-electron interaction also allows the formation of a Mott insulating state and doping of this state is theorized to drive unconventional superconductivity[8,55–58].

Around the vHSs filling, the inclusion of the local Hubbard interaction ($U_0$) and nearest neighbor site interaction ($U_1$) are proposed to induce a variety of electronic phases depending on the correlation strength compared with hopping energy. The appearance of these ordered states is very complex, and different states have been predicted in different works. From a general perspective, it is widely proposed that correlations dominated by a strong $U_0$ tend to favor magnetism (e.g. ferromagnetic, antiferromagnetic order) with strong spin fluctuations, while CDWs and superconductivity are favored by long-range correlations from a large $U_1$. The combined effect of both $U_0$ and $U_1$ can generate states like spin/charge bond orders and unconventional superconductivity[48–50]. Therefore, tuning the Fermi surfaces around vHSs can introduce different phases based on modulation of these correlation strength.

From a chemical view, the electronic properties of the kagome bands can be influenced by many factors, such as the orbital composition and bonding of the elements in kagome lattice, the neighboring out-of-plane sublattices, the orbital filling, and the electronegativity of elements[59]. Based on the analysis of bonds and band structures of many kagome materials, it was proposed that strong interlayer interaction tends to destroy the Kagome band features in materials with linked kagome lattices *i.e.* the kagome net and the nearest out-of-plane sublattice have the same or similar kinds of atoms (e.g. same group), and the bond length between the kagome atoms ($d_K$) is equal or similar with the bond length to their out-of-plane neighbor atoms ($d_{NN}$). Clear kagome bands normally appear in materials with isolated kagome lattices ($d_K/d_{NN} \lesssim 1.1$), such as CoSn, FeSn, $A$V$_3$Sb$_5$, "166" compounds as well as in semi-interacting kagome lattices ($1.1 \lesssim d_K/d_{NN} \lesssim 1.4$) such as Co$_3$Sn$_2$S$_2$. In these systems, partially-filled atomic orbitals of the atoms comprising the kagome lattice is necessary to form bonds between the kagome atoms and induce clean kagome bands near Fermi level. If the orbital is fully filled, the kagome bands tend to lie below the Fermi level and more dispersive bands arising from the interlayer bonding will dominate the electronic properties. In compounds with strongly-interacting kagome lattices ($d_K/d_{NN} \gtrsim 1.4$), most materials normally do not have typical kagome bands, such as materials with kagome lattices formed from a transition metal surrounded by a electronegative nonmetals like Cl, N, O etc (*d* electrons in kagome lattice are localized on the atomic sites), whereas, the kagome bands can exit in some compounds with kagome lattices formed by Cu, Ag, or Au atoms (e.g. Herbertsmithite, ZnCu$_3$(OH)$_6$Cl$_{12}$).

**Intertwined phases in the topological kagome metal AV$_3$Sb$_5$**

AV$_3$Sb$_5$ (A=K, Cs, Rb) is a representative kagome family that presents diverse electronic phases including a cascade of varying charge orders and a superconducting state with decreasing temperature. In this section, the studies of the origin of these states, their symmetry breaking, intertwining and competition with superconductivity (including its nature) will be discussed in detail.

*Crystalline and electronic band structures*

The family members of AV$_3$Sb$_5$ crystallize in a hexagonal structure space group *P6/mmm* (preserving inversion symmetry) and have an A-Sb2-VSb1-Sb2-A stacking order[24]. The 2D kagome net of V-atoms is located in the VSb1 layer with Sb1 atoms filling in the hexagonal centers (Figure 3a). The honeycomb lattices of Sb2 atoms are located above and below the triangle centers of kagome lattice, and the Sb2-VSb1-Sb2 sublattice is separated by the alkali atoms. Due to the weak bonds between the alkali atom and the Sb layer, the AV$_3$Sb$_5$ family is easily cleaved between the alkali layer and the Sb layer; the cleaved Sb surface is more stable than alkali surface where the alkali ions easily migrate[28].

In momentum space, the AV$_3$Sb$_5$ family presents a quasi-2D Fermi surface around the center (Γ point) of the Brillouin zone, generated primarily by its Sb $p_z$ orbitals, consistent with the strong anisotropy of the layered structure (Figure 3c). The normal state of the AV$_3$Sb$_5$ compounds is calculated to be a $Z_2$ topological metal, and the vanadium kagome net contributes to the appearance of multiple Dirac points at K and H points (Figure 3d) with gaps opening under the influence of spin-orbit coupling and topological surface states appearing at the $\bar{M}$ point near the Fermi level[60–62]. In addition, two kinds of vHSs (Figure 3e) with saddle point behavior arising from the V *3d* orbital at the M point have been observed through high-resolution angle-resolved photoemission spectroscopy (ARPES) (CsV$_3$Sb$_5$ [62,63]). One is an m-type vHS with a mixed sublattice and perfect Fermi surface nesting, the other is p-type vHS with a single sublattice. Besides the conventional vHS with quadratic energy-momentum dispersion, a p-type vHS (vHS1) near the Fermi level is found to be a higher-order vHS with energy-momentum dispersion close to a quartic polynomial along one direction, which has less pronounced Fermi surface nesting. In addition, a flat band runs across the entire momentum space below the fermi level ($E\sim-1.2$eV) alongside another nondispersive feature near the Fermi level in CsV$_3$Sb$_5$, indicating electron correlations induced by kagome lattice in the system[60].

*Charge density waves and symmetry breaking*

The AV$_3$Sb$_5$ family behaves like good metals with rich phase transitions when decreasing temperature. A first phase transition can be seen between 78-103 K (Figure 4c) and has been observed for all family members via resistivity, magnetization, and heat-capacity measurements [30,31,64,65]. This transition has been found to be a CDW onset of first-order type[66–68]. Below $T_{CDW}$, an in-plane $2a_0 \times 2a_0$ superstructure (wave vector of $Q_{3q\text{-}2a}=1/2Q_{Bragg}$) along all three directions of the kagome lattice are observed on both the Sb and alkali surfaces[28,29,69–73]. Besides the $2a_0 \times 2a_0$ CDW structure, a unidirectional $4a_0$ stripe CDW order ($Q_{1q\text{-}4a}=1/4Q_{Bragg}$) is observed in some of the STM studies of CsV$_3$Sb$_5$ (Figure 4b) and Sn-doped CsV$_3$Sb$_5$ on the Sb surface below 40 K-50 K[28,74,75]. It becomes weak with more surrounding Cs atoms on the Sb surface [76], and was not observed on Cs surface[73] nor in KV$_3$Sb$_5$[69,70]. The origin of the $4a_0$ CDW order is proposed to be a possible surface state that may be sensitive to the Fermi surface and chemical potential[72]; however its definitive origin and why it varies in the different materials and surfaces remains an open question.

Meanwhile the $C_2$ symmetry of the electronic state observed in early STM studies of the Cs surface is found to born from an electron-driven bulk nematic phase transition with $T_{nem} \sim 35$ K, demonstrated by nuclear magnetic resonance (NMR) measurement and elastoresistance measurements[73]. Coherent quasiparticles with unidirectional electronic scattering is also observed below 30 K without observation of $4a_0$ order[74]. Although the stripe-like $4a_0$ CDW order (smectic order) and nematic orders are reported in different studies, the rotation symmetry breaking from $C_6$ to $C_2$ symmetry is definitively observed in the $2a_0 \times 2a_0$ CDW structure of all of the family members. Differential conductance measurements from STM show that the energy modulation of the $2a_0 \times 2a_0$ CDW intensity along one direction is different compared with the other two directions at low temperature, thereby clearly breaking $C_6$ rotational symmetry[28,70,72,73]. Many following studies have reported that the rotation symmetry breaking and nematicity from the 3D CDW's structural distortion starts just below $T_{CDW}$[68,73,74,77–81]. In addition, three nematic domains oriented 120° to each other have been detected below $T_{CDW}$[78], as shown in Figure. 4i. The $C_2$ structural distortion is proposed to arise from a $\pi$-phase shift of the CDW between two adjacent kagome layers[73,76–78,80,82,83]. The 3D CDW phase with $2\times2\times2$ superlattice is widely detected in this family[68,76,84–86], but the $2\times2\times4$ order at l5 K[87], the $2\times2\times1$[88] and $2\times2\times4$ CDWs competition or coexistence with the $2\times2\times2$ superlattice[89], and transition from the $2\times2\times4$ (between ~60K-$T_{CDW}$) to the $2\times2\times2$ orders (below 60 K)[86] are also reported in KV$_3$Sb$_5$ via X-ray diffraction measurement. A unidirectional $\sqrt{3}a_0$ in-plane charge order is also observed in RbV$_3$Sb$_5$ that is sensitive to the surface Rb atoms desorption, implying a sensitivity to A-site stoichiometry[90]. The appearance of different orders is sensitive to temperature [86,89–91], pressure[88], surface construction and others, implying they might be close in energy and metastable under certain conditions[86,92].

The mechanism for the formation of the various CDW orderings are quite complex in the AV$_3$Sb$_5$ family. In general, the CDW order can originate from weak coupling based on Fermi surface nesting related Fermi surface instabilities[93], and also strong coupling based on electron-phonon or electron-electron interaction[94]. For AV$_3$Sb$_5$, high-resolution ARPES measurements have shown the 3D CDW induced Fermi surface reconstruction and associated band splitting below $T_{CDW}$[95,96]. Both the m-type vHS with strong Fermi surface nesting and the higher-order p-type vHSs near the Fermi level are observed to open a gap in CsV$_3$Sb$_5$[62,97–99], and the gap opening of the vHSs are observed in the original and reconstructed Fermi surfaces, as well as in KV$_3$Sb$_5$ and RbV$_3$Sb$_5$[95,100]. These results suggest the importance of Fermi surface nesting around the vHS for the formation of CDW order[62,95,97–99], which is also supported by optical spectroscopy studies[101,102]. In addition, the hard X-ray scattering in (Cs, Rb)V$_3$Sb$_5$ reported the absence of soft phonon modes, suggesting weak electron-phonon coupling[84]. A possible weak first-order transition of the CDW is also proposed, which might be related to the absence of phonon softening[82,84]. On the other hand, density functional theory calculations show unstable phonon modes at both *M* and *L* points of the hexagonal Brillouin zone, and the condensation of these phonon modes are proposed to drive the formation of the CDW[85,103–105]. Electron-phonon interaction is also reported to play an important role for the CDW state in ARPES[106] and optical study[107] of KV$_3$Sb$_5$ as well as neutron scattering experiments in CsV$_3$Sb$_5$[108]. Moreover, the Raman scattering measurement of CsV$_3$Sb$_5$ detected amplitude modes with large frequencies and strong hybridization with other lattice modes, suggesting the strong electron-phonon coupling CDW although soft phonons are absent[109]. Since electron-electron interaction is weak in AV$_3$Sb$_5$ family[106], the electron-phonon coupling and Fermi surface nesting should play important roles in the formation of CDW states.

Different ground state structures of the CDW state have been proposed in theory, but the exact CDW structure is still under debate, due to the complex charge orders and symmetry breaking. Based on an in-plane 2×2 CDW distortion without including rotation breaking and time reversal symmetry, the Star of David (SoD) and Tri-Hexagonal (TrH) structures (also called inverse Star of David in many works) are proposed candidates for the CDW phases[110]. The schematic of these structures are shown in Figure 3f and 3g. As discussed above, the CDW order is 3D, with rotation symmetry breaking that could be induced by a shift between adjacent layers. Through ARPES measurement, staggered SoD or staggered TrH structures with interlayer shifts are proposed for $KV_3Sb_5$, $RbV_3Sb_5$, and Sn-doped $CsV_3Sb_5$[95]. Both TrH ordering with interlayer shift[85,111] and alternating SoD and TrH orders[95,112] are proposed to be favored structures in $CsV_3Sb_5$, but the large number of quantum oscillation frequencies in $CsV_3Sb_5$ suggests a complex structure distortion which cannot be explained with only SoD or TrH order[113]. Inverse SoD with π-shifted along $c$ axis is also reported in $RbV_3Sb_5$[114].

Besides rotation symmetry breaking, time-reversal symmetry breaking below the CDW transition is one crucial question that is highly discussed in $AV_3Sb_5$. So far, no long range magnetic order has been observed in the $AV_3Sb_5$ family[24]. The absence of local magnetic moments was reported in an early muon spin spectroscopy (μSR) of $KV_3Sb_5$[115], and a similar result was reported in STM studies of $CsV_3Sb_5$ and $KV_3Sb_5$ with the CDW showing no magnetic field dependence[70,116], and a polar optical Kerr measurement suggesting no Kerr signal from time-reversal breaking in $CsV_3Sb_5$[117]. However on the contrary, many studies show a signal of time-reversal breaking below the CDW transition. A chiral CDW order was detected in several STM studies[69,71,72] (Figure 4e), and a clear signal of local magnetic field appearing below the CDW transition was observed in zero-field μSR measurement[118–120] (Figure 4f), indicating the time-reversal symmetry breaking (TRSB). The μSR experiment also shows an anisotropic internal field with local field direction changing from the $ab$ plane to the $c$ axis below ~30 K[120], which is consistent with the $4a_0$ or nematic order transition temperature. Additionally, electrical transport measurements detected a giant anomalous Hall effect dominated by an enhanced skew scattering effect[32,33,121], and a large anomalous thermal Hall and Nernst effect in the CDW state of $AV_3Sb_5$[122–124], consistent with the breaking of time-reversal symmetry. The signal of zero-field Kerr effect as well as circular dichroism are detected concomitant with the CDW transition[78,79,125], and nematic domains (domain size is in order magnitude of hundred μm) with opposite signs are detected without applying external magnetic field[78,125] and are magnetic field tunable[125], providing even further strong support of spontaneous TRSB.

Theoretically, the TRSB chiral flux phase[126] and charge bond order (CBO)[101,127] with orbital currents are proposed by including the nearest-neighbor ($U_1$) interaction based on the unique sublattice structures of kagome lattice. The examples of these orders are shown in Figure. 3h and 3i. The chiral flux phase with orbital loop current is proposed to be the favored state for TRSB[101], where the opposite chiral flux generated by loop current for the hexagon and triangles in the kagome lattice could induce uncompensated net flux in the unit cell. The loop current structure could explain many experimental observations such as the appearance of local magnetic field and non-zero Kerr signal [78,118,120,125]. In particular, although the $AV_3Sb_5$ family seems centrosymmetric, the second-harmonic Hall measurement shows signal of magnetochiral anisotropy for in-plane magnetic field below ~35 K[128], and the chirality is switchable by a small out-of-plane field component that can be attributed to the chiral CDW phase with orbital loop current, as shown in Figure. 4h.

*Superconductivity, PDW phase, and order intertwining*

Cooling down further to very low temperature, a superconducting state appears in the $AV_3Sb_5$ compounds with $T_c$~2.5 K in $CsV_3Sb_5$ and $T_c$~0.92-0.93 K in $KV_3Sb_5$ and $RbV_3Sb_5$ at ambient pressure[30,31,65]. The $2a_0×2a_0$ CDW phase and rotation symmetry breaking and either a $4a_0$ CDW or nematic phase is also observed to persist into the superconducting state with twofold symmetry, indicating smectic or nematic superconductivity[28,29,73,77]. In addition to these phases, a special 3Q PDW ($Q_{3q-4/3a}$=3/4$Q_{Bragg}$) phase has been detected via STM study of $CsV_3Sb_5$ (Figure 4b). Distinct from other long range charge orders, the PDW phase is a short-range order[29]. Both CDW and PDW order still exist after breaking superconductivity using a high magnetic field or raising temperature just above $T_c$. In addition, the temperature dependent d$I$/d$V$ spectrum reveals a pseudogap phase with PDW order observed only in the energy range of the pseudogap, which is different compared with the CDW phases that persist at all energies. This special 3D PDW phase is proposed to be the "mother" phase responsible for the pseudogap, in direct analogy to the cuprate high-temperature superconductors[1,29], where a primary order spontaneously breaks multiple symmetries and the resultant

"daughter" phases with lower symmetries are suggested to be the melted version of the "mother" phase.

Given the coexistence of superconductivity and charge density wave order, extensive tuning approaches including pressure, chemical/electrical doping, layer dependence and strain[129–134], have been performed to study the relation between them and understand the mechanism of these phases. The superconductivity shows prominent competition with the CDW, and multiple superconducting domes have been detected. By applying pressure in a relatively low region ($P<10$ GPa), a superconducting dome with $T_c$ reaching a maximum around $P_1$~0.5 GPa was observed in $KV_3Sb_5$, while two superconducting domes with a shallow valley between them, were observed in $(Cs,Rb)V_3Sb_5$, where $P_1$ ~0.6-0.7 GPa and $P_2$~2 GPa for $CsV_3Sb_5$[130,131]; $P_1$ ~1.5 GPa, $P_2$~2.4 GPa for $RbV_3Sb_5$[65]. The superconductivity is highly suppressed around 10 GPa in these materials. However, the reentrance of superconductivity happens when applying further high pressure, and a new SC dome is observed with $P_2$~22 GPa for $KV_3Sb_5$[132,135], $P_3$~28.8 GPa for $RbV_3Sb_5$ and $P_3$~53.6 GPa for $CsV_3Sb_5$[135–138], as shown in Figure. 4d. Associated with the superconducting domes, the CDW transition temperature reduces with increasing pressure and the CDW order is completely suppressed around $P_1$ of $KV_3Sb_5$ and $P_2$ of $(Cs,Rb)V_3Sb_5$[65,130–132], while a new CDW order starts to appear when SC reduces below $P_2$ in $KV_3Sb_5$ which may correspond to a distinct CDW state[132]; the results in $CsV_3Sb_5$ is shown in Figure. 4g. In addition, the two domes appearing in $CsV_3Sb_5$ in the low pressure region are reported to be related to the appearance of a stripe-like CDW order with $4a_0$ modulation between $P_1$ and $P_2$ by nuclear magnetic resonance experiment[139]. So far, no structural phase transition is reported under pressure in these compounds[136,138] and the tuning of the SC and CDW is proposed to be related to the changing of Fermi surface instability and electron-phonon coupling. Theoretical calculations show that the pressure reduces the $c/a$ ratio, associated with slight change of the in-plane lattice parameter, but a large enhancement of the out-of-plane Sb bonding and related out-of-plane hopping. As a result, the Fermi surface was found to have a large reconstruction on the Sb $p_z$ band, with only slight changes on the $V$ bands. Beside the influence of Fermi surface nesting, the Sb $p_z$ bands are proposed to have a large contribution to the tuning of SC and reentrant of SC at high pressure[140,141]. Associated with the band reconstruction, the reentrance of SC might be related to the Lifshitz transition and enhanced electron-phonon coupling due to partial phonon softening at high pressure [136,138].

Similar competition of the CDW and SC is also detected by electrical and chemical doping, such as Cs doping[133], Nb[134,142] and Ti[143] doping for substituting V atoms, and Sn[144,145] doping for substituting Sb. The Nb or Ti or Sn substitutions act as hole doping, which induces strong suppression of the CDW accompanied with enhancement of SC. Similar hole doping results are also found in oxidized thin $CsV_3Sb_5$ flakes[146]. Theoretically, the hole doping is found to be highly orbital selective[141,145,147], which mainly dopes the Γ band contributed by the Sb-$p_z$ orbital, and higher-order vHs with weak Fermi surface nesting. The ARPES measurement in Nb[142] reported that the shift of the vHS is found to reduce the Fermi surface nesting around $E_f$, leading to reduction of the CDW gap and suppression of the CDW order, while the expansion of Sb-derived Γ electron bands and recovery of V-derived density of states are attributed to the enhancement of SC. In addition, the electron doping by injecting Cs on the surface of $CsV_3Sb_5$ is also reported to be highly orbital selective[133], where the Sb $p_z$ orbital contributed Γ band shift downward and the Fermi surface is expanded after doping, but the CDW gap around the Fermi level is reduced. The competition of CDW and SC is also realized by reducing the film thickness of $CsV_3Sb_5$ with a maximum (minimum) of $T_c$ ($T_{CDW}$) appears around 27 layers, which is attributed to a crossover from electron-phonon coupling to electronic interactions[148]. A metal to insulating transition is detected for very thin films (below 5 layers) or through protonic gate tuning[149], which could modulate the disorder and carrier density for the modulation of SC and CDW. Monolayer $AV_3Sb_5$ is proposed to possess variety of competing instabilities like doublets of CDW and s- and d-wave superconductivity[150].

*Superconducting nature*

With diverse CDWs intertwining with SC and multiple superconducting gaps[98,151,152] detected in the $AV_3Sb_5$ family, the paring of the superconducting state and the fundamental question of conventional or unconventional superconductivity has been highly debated. In general, conventional s-wave superconductivity is fully gapped with a nodeless superconducting gap. Unconventional superconductivity with a sign-changing superconducting order parameter can induce nodes in the superconducting gap[151,153]. Conventional s-wave superconductivity with fully gapped or nodeless superconductivity was reported in some works in $CsV_3Sb_5$[66,152,153], evidenced by the appearance of the Hebel-Slichter coherence peak at ambient pressure in the nuclear quadrupole resonance (NQR) and NMR measurement[66], and the exponential behavior of the magnetic penetration depth $\Delta\lambda(T)$ that deviates from a linear or power law temperature dependence for nodal superconductivity[153]. In addition, the direction characterization of superconducting gap using STM showed distinct gap shapes on distinct surfaces, which is likely dependent on the complicated Fermi surfaces. A V-shaped SC gap (possibly related to Fermi surfaces contributing to an anisotropic CDW gap) and a U-shaped SC gap (isotropic and less affected by the CDW) were seen on both the half-Cs and Sb surfaces of $CsV_3Sb_5$[151]. These types of SC gap might be sensitive to the modulation of Fermi surface[154]. The s-wave superconductivity is proposed based on the in-gap states that can be induced by magnetic clusters but not nonmagnetic impurities[151]. Also, early $\mu$SR experiments reported no time-reversal symmetry breaking in superconducting state because of no change of muon spin relaxation rate during SC transition[152,155].

However, unconventional superconducting signals have been observed in many other investigations. In another $\mu$SR experiment, they found nodal superconductivity in $RbV_3Sb_5$ and $KV_3Sb_5$ with magnetic penetration depth showing linear-in-$T$ behavior at ambient pressure, while a transition from nodal to nodeless superconductivity is observed when increasing pressure[156]. In particular, when the signal of the charge order is strongly suppressed under pressure, the zero-field muon spin relaxation rate showed a clear increase below the superconducting transition measured at 1.1 Gpa for $KV_3Sb_5$ and 1.85 Gpa for $RbV_3Sb_5$, clearly indicating time-reversal breaking in the superconducting state and implying an unconventional paring state[156]. Also, the $\frac{T_c}{\lambda_{ab}^{-2}}$ in $(K,Rb)V_3Sb_5$ shows large value ($\lambda_{ab}$ is the in-plane magnetic penetration depth) which is far from conventional superconductivity[119,156]. In $CsV_3Sb_5$, nodal superconductivity is also reported by measuring thermal conductivity[137]. The NQR experiments reported the absence of the Hebel-Slichter coherent peak when the pressure is above the $P_2$ of second superconducting dome[139] in $CsV_3Sb_5$, also suggesting unconventional superconductivity. Moreover, other STM works reported unconventional zero-bias conductance peaks in the V-shaped SC gap of the Vortex-Core state on the half-Cs surface of $CsV_3Sb_5$, which may correspond to possible Majorana bound states from possible topological surface states[76]. Superconductivity of the surface states and signals of possible spin-triplet superconductivity was also seen in proximitized $K_{1-x}V_3Sb_5$ Josephson junctions[157].

Theoretically, both s-wave and unconventional superconducting paring are proposed to be possible in the $AV_3Sb_5$ family. It was reported that the electron-phonon coupling is too weak for the superconducting $T_c$ in these materials, suggesting possible unconventional pairing[104]. In this family, the local correlation (on-site) is also proposed to be weak and nonlocal electron correlation might be crucial[158]. By including both local correlation and nearest-neighbor Coulomb repulsion, rich pairing symmetries are found near the vHSs with competitive spin-triplet pairing and spin-singlet pairing[159–161]. The favored pairing state is highly dependent on the nonlocal Coulomb interaction, interaction strength, fermi momentum and van Hove bands[159–161]. For instance, a dominant $f$-wave spin-triplet superconducting order is theorized to be favored for weak coupling while d-wave singlet pairing competes with $p$-wave pairing under very strong correlation[159]. The complex electronic bands with multiple vHSs and the sensitivity of superconductivity on the Fermi surface and electron correlations shows the fertile possibility of controlling superconducting paring in the $AV_3Sb_5$ family. Another interesting topic is the possible unusual charge pairing (e.g. charge 4e or 6e) beyond the normal charge-2e pairing superconductivity[162–165]. These unusual charge paring may happen based on the PDW superconductivity or nematic/stripe superconducting state[165–167]. Charge 4e and 6e superconductivity was recently reported in $CsV_3Sb_5$ by measuring the magnetic flux quantization during superconducting transition region based on Little-Parks effect[168]. And the effective flux area in the Little-Parks measurement[169] and non-uniform vestigial charge-4e phase are discussed in theory[170]. The investigation of these unusual charge pairing has just begun and more detailed investigations are needed to have a clear understanding of the paring mechanism.

Overall, following the extensive study of $AV_3Sb_5$ family, more and more understanding of charge orders and superconducting properties have been achieved. The clear investigation of these properties still require more effort, especially the elucidation of the charge order ground state, TRSB order state, and superconducting pairing, which are of critical importance for future further investigation of unconventional superconductivity and topological superconductivity. So far, most of the studies in $AV_3Sb_5$ are based on bulk samples; only a few investigations have been on thin samples. The thin films of the $AV_3Sb_5$ might also be beneficial to further investigation of CDW and superconductivity, making in-situ tuning of Fermi surfaces in nanodevices possible for broad investigations. In addition, future studies using local field sensitive approaches in the nanoscale would also be extremely beneficial for understanding the correlated charge orders and origin of time reversal symmetry breaking. Techniques such as nitrogen-vacancy magnetometry, magnetic force microscopy and scanning superconducting quantum interference devices[171–173] could be very helpful in this regard.

## Beyond $AV_3Sb_5$: Emerging kagome materials for future research

### Kagome metal $AM_6X_6$

Recently, another kagome metal family, $AM_6X_6$, has attracted substantial attention due to its more prevalent magnetic properties and large chemical diversity; the A site can be alkali, alkali earth and rare earth metals, the M site can be a transition metal such as Mn, Fe, Ni, V, Co, Cr, and the X site is typically a group IV element such as Sn, Ge, Si. In this family, most of the "166" compounds crystallize in centrosymmetric structures analogous to the CoSn structure type (or B35) consisting of graphite-like Sn nets and kagome nets of Co centered by Sn atoms. Depending on the relative positions of the *A* site atoms and capping of the transition metal atoms in the CoSn host structure, these intermetallics can adopt the hexagonal $HfFe_6Ge_6$-type (H-type) or the hexagonal $YCo_6Ge_6$-type (Y-type, a disordered variant of the H-type). The schematic of these structures are shown in Figure. 5a.

The H-type $HfFe_6Ge_6$[174], *P*6/*mmm* space group with cell dimensions of $a = 5.065$ Å and $c = 8.058$ Å, is formed by 'stuffing' Hf atoms into the hexagonal voids of the FeGe (CoSn-type) framework. Figure 5a shows the stacked layers of alternate filled and empty planes of the $HfFe_6Ge_6$ (H-type) crystal structure. This ordering of the Hf atoms (A site) in the alternate layers results in a doubling of the unit cell along the *c* direction, where the kagome nets are separated by inequivalent $Ge_3$ layers and $HfGe_2$ layers. Many "166" compounds like $RMn_6Sn_6$ compounds (*R*= rare earth) possess the H-type $HfFe_6Ge_6$ structure.

The Y-type, $YCo_6Ge_6$ structure[175] also crystallizes in the *P*6/*mmm* space group with unit cell dimensions of $a = 5.074$ Å, $c = 3.908$ Å. Figure. 5a shows the $YCo_6Ge_6$ structure type, formed by stuffing Y atoms into a CoGe (CoSn-type) hexagonal framework. In contrast to the order of the Hf atoms in the H-type, the Y atoms in this structure type are occupationally disordered. Despite the similarity of the H-type $HfFe_6Ge_6$ and Y-type $YCo_6Ge_6$ structures, they are distinguishable by powder diffraction by the *d* spacing of the (*001*) reflection. Materials with Y-type $YCo_6Ge_6$ structures in "166" compounds include $SmMn_6Ge_6$, $RdMn_6Ge_6$, and $RFe_6Sn_6$ (*R*=Y, Gd-Er))[176–179].

In addition to the H-type $HfFe_6Ge_6$ and Y-type $YCo_6Ge_6$ structures, some compounds also crystallize in other structures with lower symmetry, such as the $HoFe_6Sn_6$-type[177,178] (*Immm* space group, materials include $YbFe_6Ge_6$, $LuFe_6Ge_6$), $TbFe_6Sn_6$-type[177,178] (*Cmcm* space group, materials include $YFe_6Ge_6$, $TbFe_6Ge_6$, $DyFe_6Ge_6$, $HoFe_6Ge_6$,), and $GdFe_6Sn_6$-type[177] with *Pnma* space group. The large material family of "166" compounds provide plenty of opportunities to tune the magnetic interaction and electronic correlation, depending on different interactions between the various elements and lattices. A broad range of magnetic orders, metaphases, and topological states have been observed and predicted.

*Diverse magnetic states*

Magnetism has been the most intensely investigated property in these compounds over the past few decades, with diverse magnetic properties (unlike the $AV_3Sb_5$ family) observed varying from ferromagnetic, ferrimagnetic, antiferromagnetic, spin chirality, to weak or non-magnetic. The "166" compounds with a kagome layer of Fe and Mn generally possess obvious magnetic textures, e.g. $RT_6X_6$ compounds (*R*= rare earth; *T*= Mn, Fe; *X*=Ge, Sn)[36,121,177,178,180–185], in which the magnetism is influenced by the inter- or intra- layer interaction of rare earth elements and the magnetic transition metal kagome sublattices.

Keeping the transition metal in the kagome lattice constant and changing the rare earth elements can introduce great differences of the magnetic order with a typical example being the widely studied $RMn_6Sn_6$ compounds with $HfFe_6Ge_6$-type structure. $RMn_6Sn_6$ compounds with *R* being magnetic rare earth elements (*R*=Gd, Tb, Dy, Ho) possessed strong interactions of 4*f* electrons (*R* atom) and 3*d* electrons (Mn atom)[36,183–185]. These compounds have ferromagnetic layers of both *R* and *M* atoms, with ferromagnetic Mn-Mn inter-sublattice exchange, but strong antiferromagnetic Mn-*R* inter-sublattice coupling. As a result, the compounds present a collinear ferrimagnetic behavior below the curie transition temperature (between 376 K and 435 K)[184,185]. The magnetic order was reported to lay in-plane for $R$=Gd, while additional spin reorientation was also observed for $R$= Tb, Dy, Ho. In TbMn$_6$Sn$_6$[183,186], the magnetic order is along the $c$-axis at low temperature and deviates from it at room temperature. Whereas, for $R$ = Ho and Dy, a conical magnetic structure with magnetic order deviated by 45°- 50° from the $c$-axis below 180-240 K[183]. The schematic magnetic ground state of Mn and $R$ atoms in $R$Mn$_6$Sn$_6$ are shown in Figure 5d.

In $R$Mn$_6$Sn$_6$ compounds with $R$ = Er, Tm and diamagnetic rare-earth elements without 4$f$ electrons ($R$= Sc, Lu, Y), the magnetic coupling between $R$ and Mn lattices are quite weak, which presents a paramagnetic-antiferromagnetic transition with Neel temperature $T_N$ between 340 K-384 K[184,185]. In ErMn$_6$Sn$_6$ and TmMn$_6$Sn$_6$, the magnetic ordering of erbium and thulium sublattices introduces a second transition with ferrimagnetic (~75 K) and antiferromagnetic ordering (58 K)[185], respectively. However, in compounds with $R$= Sc, Lu, Y, the magnetism is mainly dominated by the alignment of magnetic order in Mn sublattices[185]. They were reported to have complex magnetic order due to interlayer magnetic interactions of nearest and second-nearest-neighbor Mn kagome lattices, which introduces a transition with incommensurate antiferromagnetic ordering. Below this transition (~333 K for YMn$_6$Sn$_6$), the Mn moments in YMn$_6$Sn$_6$ form the "double flat spiral" structure without applying magnetic field[187], where the in-plane moment of the Mn layer changes orientation with respect to each other, resulting in a spiral structure along the $c$-axis. The spiral structure was also reported in LuMn$_6$Sn$_6$, ScMn$_6$Sn$_6$, and even TmMn$_6$Sn$_6$, ErMn$_6$Sn$_6$[188–190].

In addition to $R$Mn$_6$Sn$_6$ compounds, distinct magnetic properties were reported in other compounds. Ferromagnetic orders were reported in manganese stannides by replacing the $R$ element to alkali or alkali earth elements with low concentrations of valance electron donors, such as Li, Mg or Ca. The Curie transition temperature is about 382 K for LiMn$_6$Sn$_6$ (easy plane parallel to kagome lattice) and 250 K- 290 K for compounds with Mg and Ca[191,192]. Compared with $R$Mn$_6$Sn$_6$, $R$Mn$_6$Ge$_6$ also has different magnetic behaviors, found to be ferrimagnetic for $R$= Nb, Sm (Curie temperature between 417 - 441 K)[179], but antiferromagnetic for $R$=Dy-Yb, Sc, Y, Lu, Gd (Neel temperature increase for heavier rare earth elements)[177,193]. In the iron compounds ($R$Fe$_6$Sn$_6$ and $R$Fe$_6$Ge$_6$), the strong easy $c$-axis anisotropy of Fe lattices is reported to dominate the magnetic anisotropy, and most compounds have antiferromagnetic orders below a Neel temperature $T_N$> 400 K[177,181]. Different from the Mn/Fe-based compounds, AM$_6$X$_6$ compounds with non-magnetic $B$ site variants allow the generation of non-magnetic kagome metals (e.g. YV$_6$Sn$_6$)[194] and compounds with weak magnetism (e.g. ScV$_6$Sn$_6$, GdV$_6$Sn$_6$)[194,195]. In the cobalt variants, MgCo$_6$Ge$_6$[196] is paramagnet and weak magnetism was observed in Yb$_{0.5}$Co$_3$Ge$_3$[41,42].

In compounds with complex inter/intra interactions between $R$ and Mn/Fe layers, magnetic reorientation and rich magnetic phases can also be generated by applying magnetic field. For instance, the out-of-plane magnetic field (along $c$-axis) was found to gradually tilt the magnetic moment in $R$Mn$_6$Sn$_6$ ($R$= Dy, Tb, Ho, Gd) towards the $c$-axis, while a metamagnetic transition happens in ErMn$_6$Sn$_6$[36]. In YMn$_6$Sn$_6$, it was reported that the in-plane magnetic field (aligned with the kagome plane) can change the magnetic state from the distorted spiral state, to a transverse conical spiral phase (TCS), fan-like state, (a quadrupled structure along the $c$-axis with spins deviating from in-plane magnetic field direction), and a forced ferromagnetic state with increasing magnetic field[187,197], as shown in Figure 5f and 5g. In particular, a topological Hall effect was reported in YMn$_6$Sn$_6$[198] and TmMn$_6$Sn$_6$[199] without a field-driven skyrmion lattice, and the microscopic origin of this is proposed to originate from dynamic chiral fluctuation driven non-zero spin chirality in TCS phase[187,198–200].

*Topological states and flat band*

Based on the 3d-transition metal kagome lattice, "166" compounds can host topological flat bands, van hove singularities, and Dirac points in their band structure. In particular, the diverse magnetic coupling in these kagome magnets provides new opportunities to modulate the electronic bands and introduce novel topological states and quantum phases.

Very recently, the detection of all of these characters was reported through ARPES measurement in the GdV$_6$Sn$_6$[201], HoV$_6$Sn$_6$[202], YCr$_6$Ge$_6$[203], and YMn$_6$Sn$_6$[204]. In addition, topological surface states were also detected in GdV$_6$Sn$_6$ and HoV$_6$Sn$_6$[201,202], owing to the ideal kagome lattice and large bulk gap around Γ that allows well separated (or unmixed) topological surface states from bulk states. The detection of topological surface states in these materials is distinct on different surface terminations, and is sensitive to chemical doping on the surface.

For YMn$_6$Sn$_6$, the vHSs, Dirac cone and flat band are observed to be quite close to the Fermi level[204]. In particular, YMn$_6$Sn$_6$ has in-plane ferromagnetic ordering of its Mn layer but helical antiferromagnetic ordering along the *c*-axis. Different with the paramagnetic state which normally has degenerate bulk bands, the spin degeneracy is theorized to be lifted in the magnetic state of YMn$_6$Sn$_6$ due to the ferromagnetic state with large spin orbit coupling. Spin-polarized APRES measurements are needed to confirm the band structure of magnetic YMn$_6$Sn$_6$.

In particular, kagome materials when possessing strong intrinsic SOC and out-of-plane ferromagnetism, have the ability to realize nontrivial Chern phases, in which a gap is opened in the spin-polarized Dirac bands with spinless topological edge states lying in the gap[54,205–207] (Figure. 5b). Taking advantage of the pure manganese kagome layers and a strong out-of-plane ferromagnetic order as well as a defect-free kagome lattice, a quantum-limited Chern phase was reported in TbMn$_6$Sn$_6$ using STM measurement (Figure. 5e)[35]. The magnetic Laudau quantization, as well as localized edge states, in the Chern gap were detected. In addition, TbMn$_6$Sn$_6$ presents an anomalous Hall effect with a prominent intrinsic contribution speculated to arise from the Berry curvature generated from Chern gap Dirac fermions or other complex Dirac bands[35,208].

Similar to TbMn$_6$Sn$_6$, Chern gapped Dirac fermions were reported to exist in other *R*Mn$_6$Sn$_6$ compounds with ferromagnetic kagome lattices, e.g. *R*=Gd-Er, and also proposed in YMn$_6$Sn$_6$[204]. In *R*Mn$_6$Sn$_6$ compounds (*R*=Gd-Er)[36], the existence of Chern gapped Dirac fermions was studied based on transport measurements of quantum oscillations with nontrivial Berry phases, as well as the large intrinsic anomalous Hall effect. In addition, the Dirac cone energy and gap size were reported to gradually decrease when *R* was changed from Gd to Er[36]. This highlights the possibility of Chern gap tuning based on tuning magnetic exchange coupling between the rare earth element with 4*f* electrons and the transition metal with 3*d* electrons. In addition to these compounds, introducing new "high entropy" phases by mixing multiple elements in the *R*Mn$_6$Sn$_6$ system, e.g. (Gd$_{0.38}$Tb$_{0.27}$Dy$_{0.2}$Ho$_{0.15}$)Mn$_6$Sn$_6$, can generate new magnetic transitions based on the competing magnetic interactions of the various elements at the mixed site[209]. It was reported that (Gd$_{0.38}$Tb$_{0.27}$Dy$_{0.2}$Ho$_{0.15}$)Mn$_6$Sn$_6$ presents similar magnetic transitions as TbMn$_6$Sn$_6$ at high temperature, but new ferrimagnetic transitions with tilted magnetic moments (30º) around 90 K which were not seen in *R*Mn$_6$Sn$_6$ (*R*=Gd, Tb, Dy, Ho). The high entropy phase also shows an intrinsic AHE effect, suggesting possible Chern gapped Dirac fermions[209].

*Charge density wave and structural transition*

Although some of the kagome magnets discussed above have strong correlations with saddle points and flat bands in electronic bands, *no charge density wave has been observed* in those systems to date. On the contrary, the CDW state and structural transitions were reported in several other "166" compounds with specifically weak magnetism or a nonmagnetic kagome lattice, including ScV$_6$Sn$_6$[195] (HfFe$_6$Ge$_6$-type), MgCo$_6$Ge$_6$ (HfFe$_6$Ge$_6$-type)[196], also in Yb$_{0.5}$Co$_3$Ge$_3$[41,42]-a hybrid structure of YCo$_6$Ge$_6$ and CoSn. This is similar to the case of the AV$_3$Sb$_5$ family discussed above which also has weak or nonmagnetic kagome lattices, although the details of the transitions and ground states are different, suggesting a competition between the correlated versus magnetic ground states generally exists in Kagome lattices.

Similar to the AV$_3$Sb$_5$ family, ScV$_6$Sn$_6$ hosts a vanadium kagome lattice with a partly filled *d*-orbital. A transition signal around 92 K was observed in resistivity, magnetization and heat capacity, which is confirmed to be a CDW transition by single-crystal X-ray diffraction [195]. Different from AV$_3$Sb$_5$, hosting 2x2x2 or 2x2x4 CDW orders, the distorted lattices in ScV$_6$Sn$_6$ was reported to possess 3x3x3 CDW orders. In addition, the ScV$_6$Sn$_6$ shows strong modulated displacement of the Sc and Sn atoms in CDW state, but very weak displacements of the vanadium atoms, distinct from the V lattice displacement dominated lattice rearrangement found in AV$_3$Sb$_5$. The partly filled *d*-orbital kagome bands were proposed to be related to the CDW and more studies are needed to elucidate the detailed CDW order state, origin of CDW transition and the relation to V-based kagome bands in ScV$_6$Sn$_6$.

Compared with AV$_3$Sb$_5$ and ScV$_6$Sn$_6$, the structural transition observed in the Co-based kagome compounds is quite different. Yb$_{0.5}$Co$_3$Ge$_3$ crystallizes in *P*6/*mmm* space group with one kagome layer in a unit cell at room temperature[41]. A phase transition associated with a resistivity anomaly was observed in Yb$_{0.5}$Co$_3$Ge$_3$ around 95 K. Below this transition, the Yb and Ge atoms retain their geometries, while the Co atoms distort the kagome plane by "twisting" the triangles with opposite distortion direction between two neighboring kagome lattices, resulting in a doubling of the unit cell along the *c*-axis, crystallizing in the *P*6$_3$/*m* space group that breaks $C_6$ rotation symmetry[42]. MgCo$_6$Ge$_6$ also has similar transitions with twisting of the kagome lattice at low temperature[210], and the structure changes from the *P*6/*mmm* (HfFe$_6$Ge$_6$-type) space group at room temperature to *P*6$_3$/*mcm* at 100 K[196]. This twisted structure was proposed to be related to the splitting of 3*d* orbital in Co and interaction of Co-Ge atoms. Unlike AV$_3$Sb$_5$ and ScV$_6$Sn$_6$, which have clear magnetization variation commensurate with the structural transition, there is no change of magnetization during the structural transition in Yb$_{0.5}$Co$_3$Ge$_3$ and MgCo$_6$Ge$_6$. This unusual feature has not been fully understood, but it may be related with the particular twisting of Co lattices. And while a CDW transition without related magnetic transition has been reported in other kinds of correlated materials (e.g.NdNiC$_2$)[211,212], Yb$_{0.5}$Co$_3$Ge$_3$ shows a weak magnetic transition signal around 18-25 K, along with the appearance of negative magnetoresistance below the transition temperature[42], meaning there is further unresolved complexity in the low temperature state than what is currently understood. It's also worth noting some materials in RT$_3$X$_2$ (R=rare earth, T=transition metal, X=Si, Ga, B) kagome family (132 family) with similar twisted structures are primarily observed to be superconductors, but without a structural transition from room to low temperature. These include LaRu$_3$Si$_2$, LaIr$_3$Ga$_2$, YRu$_3$Si$_2$[213–216]. It may be very interesting to explore whether the superconductivity can appear in Co based 166 compounds with their structural transition, or if it arises upon suppression of the transition. Possible chemical doping and high pressure approaches can be used to explore the structural and magnetic phase transitions and possible superconductivity in Yb$_{0.5}$Co$_3$Ge$_3$ and MgCo$_6$Ge$_6$.

Overall, the "166" compounds with different kagome lattices and interlayer interactions are a great platform to understand the interplay between CDW orders and magnetic properties. There is ample opportunity to investigate distinct spin related physical properties such non-trivial topological states, topological Hall, and superconductivity – and its interplay with magnetic and topological states.

**The Trigonally Distorted (a.k.a breathing mode) kagome system (Nb$_3$X$_8$, X = Cl, Br, I)**

One highly interesting structural modulation of the ideal kagome lattice is the trigonal distortion, also called the "breathing" kagome lattice, described previously and shown in Figure 2d. An important example material with the breathing kagome lattice is Nb$_3$X$_8$ (X = Cl, Br, I)[38,217], possessing different physical properties compared with the AV$_3$Sb$_5$ (A = K, Rb, Cs) and '166' materials aforementioned.

The Nb$_3$X$_8$ family is also a vdW type material and has extremely exfoliatable layered structures, more easily achieving ultrathin flakes than the AV$_3$Sb$_5$ family for example, as individual layers are very weakly coupled through vdW interactions[37,218]. Nb$_3$Cl$_8$, Nb$_3$Br$_8$ and their intermediate phase Nb$_3$Cl$_{8-x}$Br$_x$ have two different crystal structures, corresponding to high-temperature α phase with space group of $P\bar{3}m1$, and a low-temperature β phase with space group of $R\bar{3}m$[37,219,220], as shown in Figure. 6b. Both phases have the same single layer structure, but can be stacking with different orders, hence the different unit cells. A single layer of Nb$_3$X$_8$ shows the breathing kagome structure formed by the Nb atoms, where three Nb atoms are trimerized through strong metal-metal bonds as shown in Figure. 6a. Each Nb atom is under a distorted octahedral environment of X atoms, and each layer has a sheet of Nb atoms that is sandwiched between two sheets of X atoms. In addition, the two sheets of X atoms are not equivalent, resulting in a lack of inversion symmetry (a.k.a. noncentrosymmetric) in a single layer. The α phase contains a unit cell of two layers with the second layer being an inversion of the first layer. The β phase contains six layers with each layer resulting from the inversion operation of the previous layer with a translation, hence glide planes are present. In addition, both the α and β phases have inversion centers located at every vdW gap, and as a result even-layered Nb$_3$X$_8$ preserves inversion symmetry while odd-layered Nb$_3$X$_8$ breaks it. Finally Nb$_3$I$_8$ shows the same crystal structure as Nb$_3$Br$_8$ (β phase) at room temperature but there are no reports of a transition as yet; it's likely that it will undergo the same transition with Nb$_3$Cl$_8$ and Nb$_3$Br$_8$ at higher temperature, but more study is necessary.

*Magnetic properties*

Nb$_3$Cl$_8$ and Nb$_3$Br$_8$ in both bulk and powder forms undergo a phase transition from a paramagnetic state to a nonmagnetic singlet state[37], accompanying the structural transition from α to β phase[37]. The magnetic susceptibility vs. temperature measurements reveal that the transition temperature of $Nb_3Cl_8$ is ~90 K and increases when Cl is substituted by Br, and reaches ~380 K for $Nb_3Br_8$[37], as shown in Figure. 6c. It's worth noting that at room temperature, bulk $Nb_3Cl_8$ is paramagnetic with a two-layer unit cell, whereas at bulk $Nb_3Br_8$ is nonmagnetic with a six-layer unit cell, but they share the same β phase at low temperatures. Another feature in Figure 6c is that there are upturned tails when cooling the samples down to around 10-20 K; it was proposed that these originate from defect spins induced by broken singlets or trapped high temperature phases during synthesis since the tails are sample dependent rather than being material intrinsic. Though there are no magnetic experimental studies on $Nb_3I_8$ to date, either in bulk or powder, from theoretical calculations, it's expected that it has the same magnetic transition[218,221], at a higher temperature.

One open question arising from the magnetic properties that have not been characterized down to 2D limit, is whether the transition temperature changes with the thickness as the flakes goes thin, and whether the few layered flakes possess different the magnetic properties than the bulk. Theoretical calculations predict that the ground state of $Nb_3X_8$ in monolayer form is ferromagnetic[221,222]. In $Nb_3I_8$, some work also predicted that it has layer dependent magnetism; for monolayer and bulk, it is ferromagnetic, but for bilayer and trilayer, it is antiferromagnetic due to interlayer coupling. In addition, different theoretical calculations yield very different Curie temperatures for monolayer $Nb_3I_8$; 307.5 K in Ref. [221][221] vs 87 K in Ref. [222][222]. Moreover, the current theoretical calculations for the $Nb_3X_8$ family only used the Ising spin model (easy axis of spin orientation is along the z-axis)[221,222]; the Heisenberg model which also includes in-plane spins (magnetic anisotropy, which is very important in 2D materials) can be considered to improve the calculation. Since the parent kagome lattice hosts geometric frustration and the breathing lattice maintains the triangular geometry, these kinds of interactions also need to be considered while calculating the magnetic ground state of $Nb_3X_8$. Given the discrepancies of the properties predicted, more detailed studies on the magnetization of bulk and few layer $Nb_3X_8$, including μSR as was done for the $AV_3Sb_5$ family, are required to elucidate the question of their ground magnetic state.

*Electronic properties: Correlations and Obstructed Atomic Insulating states*

Most *ab initio* theoretical calculations will show a metallic band structure of $Nb_3Cl_8$ and $Nb_3Br_8$ in their bulk form[39,40,220,223]. However, in both transport and ARPES measurements, they are proven to be insulating[40,223,224] with band gaps on the order of an eV. Again such a large discrepancy between theoretical and experimental results indicates the presence of the expected nontrivial effects from the kagome lattice. One possible explanation of this abnormal insulating state is that $Nb_3Cl_8$ and $Nb_3Br_8$ may have very strong correlations. They are predicted to be Mott insulators with the kagome flat band lying near the Fermi level being split into two bands, the so-called upper and lower Hubbard bands, yielding the observed insulating state[39,40]. Such an explanation agrees well with recent ARPES measurement in $Nb_3Cl_8$ bulk[40].

Another possible explanation is the obstructed atomic insulator theory[225,226], which predicted that in the β phase of $Nb_3Br_8$ and $Nb_3I_8$, there are centers of charges not localized on the atoms but sitting in the vdW gaps in every other layer. Therefore, while $Nb_3Br_8$ and $Nb_3I_8$ flakes with an odd number of layers can be metallic, flakes with an even number of layers can be either metallic or insulating based on the different termination conditions. If the charge centers are exposed on the surface as a trivial surface state the flakes can be metallic, but if the charge centers are buried inside the vdW gap, they will be insulating.

In the absence of strong correlation and obstructed atomic insulator theory, both theoretical calculations and ARPES experiments showed an insulating state at room temperature in $Nb_3I_8$ (β phase)[221]. The band structure is also predicted to be thickness dependent, being metallic in the monolayer, insulating as a bilayer, metallic as a trilayer, with corresponding thickness dependent magnetic properties[221].

It's necessary to mention that many of these electronic band structure results are calculated without including spin-orbit coupling. The theoretical calculations that do, also predicted that upon introducing spin-orbit coupling, the monolayer of the $Nb_3X_8$ family will show a spontaneous valley polarization as large as ~100 meV[227]. Among this family, $Nb_3I_8$ is proposed it to be a good platform for investigating a valley Hall effect. $Nb_3Cl_8$ and $Nb_3Br_8$ are less ideal because their spin-down bands are predicted to be located in the gap of the spin-up bands.

*Electronic Properties: Flat band*

Although there are different predications of the electronic band structures, a common feature shown in many studies is there is still the prominent flat band in $Nb_3X_8$ family. As introduced before, flat bands exist in the ideal kagome lattice, however, it's extremely isolated in the trigonally distorted lattice. Recent ARPES measurements showed the flat bands below the Fermi level in $Nb_3Cl_8$ bulk (Figure. 6d), and $Nb_3I_8$ bulk (Figure. 6e) at room temperature[228]. For instance, the flat band in $Nb_3Cl_8$ bulk lies at around 0.72 eV below the Fermi level with a bandwidth of 0.28 eV[40]. Note that, the crystal structure of $Nb_3Cl_8$ is α phase whereas $Nb_3I_8$ is β phase at room temperature, having a different glide plane symmetry between layers, indicating that the flat band is robust against the layer stacking situation, likely due to the weak vdW interaction between layers.

Though the ARPES results confirmed the presence of the flat band in these materials, it has not yet been studied via transport due to their highly insulating state, making accessing it much more difficult. There are at least two alternative ways to overcome this obstacle; first, by making thinner $Nb_3X_8$ flakes as it has been theorized that at least monolayer $Nb_3X_8$ will be metallic while preserving its flat band and simultaneously pushing it very close to the Fermi level. Second, though bilayer and thicker layer $Nb_3X_8$ might still be insulating, they should also be gate tunable, which allows the different filling states of the flat band to be investigated. And while the $Nb_3X_8$ family shows intriguing band structures and abnormal magnetic properties in bulk, the studies of their layer dependent properties are still very few. They showed a unique property when integrated into heterostructures, e.g., $Nb_3Br_8$ was adopted as a barrier layer in a Josephson junction[229], forming a $NbSe_2/Nb_3Br_8/NbSe_2$ heterostructure, and showed the field-free Josephson diode effect (non-reciprocal superconductivity). These observations indicate that $Nb_3X_8$ family houses more exotic physical properties, especially in their few layer forms, and their 2D exfoliability and integratabily into heterostructures make them an ideal platform to study the isolated kagome flat band, Mott insulating state, and the trigonal distortion.. Additionally, manufactured correlation through the formation of moire superlattices via twisted stacking is another important knob that can be tuned in this family. Their study is an extremely fertile area for deeply diving into correlated physics.

**Conclusion and perspective**

Recent studies have shown many kagome materials with different kinds of features, from metallic to insulating, non-magnetic to hosting complex magnetic textures, and relatively weak electron correlation to very strong correlation (Table 1). Based on the combination, interplay, and competition of topology, electron correlation, and magnetism, kagome systems provide broad opportunities for investigating and understanding exotic physical phenomena (Figure. 7) and the entanglement of quantum phases in future research.

One of these directions is the intertwining and tuning of different ordered states. The appearance of ordered states in materials is very complex and clear conclusions are difficult to obtain since they are very sensitive to the electronic state around Fermi level. Nonetheless, experimental and theoretical studies suggest some clues in understanding the various ordered states. As discussed in the second part of this review, around the van Hove filling, different electronic orders can appear depending on the strength of different electron interactions. The intertwining and competition of many ordered states (e.g. SC state, liquid crystal phases) appear in materials with relatively weak or intermediate electron correlation strength, such as in $AV_3Sb_5$. For $AV_3Sb_5$, although the relation of electronic orders and superconductivity have been studied extensively, the explicit charge ordered states breaking time reversal symmetry and the nature of its superconductivity remains elusive, which is a challenging but essential topic for further research. Another other phenomenon is the formation of charge instabilities induced states (e.g. CDW) in kagome materials that are mostly observed in materials with weak magnetism, as seen in the "166" compounds and $AV_3Sb_5$. Recently, a short-range CDW order was also reported in the antiferromagnetic state of correlated kagome metal FeGe[230], which is proposed to arise from vHSs driven Fermi surface instability near the Fermi level. The coupling of magnetic orders and electron correlation is another important direction to explore, e.g. understanding the relation of CDW and magnetic order (their competition but also how they combine to influence, for example, superconductivity), as well as investigating SDWs in the magnetic states of materials with strong Fermi surface nesting around Fermi level.

Another direction is investigating the coupling of topology with other orders, which allow the exploration of unconventional phases. The past combination of topology and magnetism has shown the potential to induce new topological states in magnetic kagome materials, e.g. Chern quantum phases $R$Mn$_6$Sn$_6$ family with strong spin-orbit coupling, associated with large Berry curvature induced strong anomalous Hall effect, anomalous thermal Hall[180,231]. These are just the beginning; topological kagome magnets have a large opportunity for broad investigations including the quantum anomalous Hall effect, Weyl states, charge-spin conversion and spin torque based spintronic devices utilizing spin polarized topological bands, and skyrmion formation from exchange coupling tuning. The $AM_6X_6$ family's diverse magnetism is a good platform for investigating the relation of topological states and magnetic coupling. The $AV_3Sb_5$ family is good for investing weak to intermediate correlation strength with topological states and the $Nb_3X_8$ family offers a good platform for understanding strong correlation with topology and thickness/size dependence. Moreover, inducing superconductivity in topological states, either intrinsically or through the proximity effect, are promising for exploring unconventional and topological superconductivity. In fact, many compounds in $RT_3X_2$ (R=rare earth, T=transition metal, X=Si, Ga, B)[232,233] possess intrinsic superconductivity, and properties including flat band, Dirac cones, and vHS are recently being proposed[214,234]. $RT_3X_2$ will be another interesting family for investigating the interplay of superconductivity, topology, magnetism and correlation physics.

With more and more clear flat bands being observed in many kagome materials, the flat band associated quantum phases will be a critical focus in related area. As discussed before, partially filling of flat band could induce quantum phases such as Mott-insulating state, nontrivial Chern topology, unconventional superconductivity, and a fractional quantum Hall effect. Materials with isolated flat bands (e.g. $Nb_3X_8$), are of great potential to realize these states, by modulating the filling of bands and engineering of the band structure through chemical doping, electrostatic tunning or mechanical modification. In addition, kagome antiferromagnets with spin-orbit coupled Mott insulating state might be candidates for investigating quantum spin liquid state based on frustration of strongly localized magnetic moments. Finally, leveraging the multitude of exotic phenomena these materials possess, makes them very interesting for integration into nanoscale electronic devices. Hence kagome materials are extremely important and exciting quantum materials and will be a thriving area of investigation for a very long time.

**Acknowledgements**

Y.W. acknowledges the support from NWO Talent Programme Veni financed by the Dutch Research Council (NWO), project No. VI.Veni.212.146. H.W. acknowledges support from the Kavli Institute of Nanoscience Delft Synergy Grant 2022. Y.W., H.W., and M.N.A. acknowledges support from the Technical University of Delft Quantum Nanoscience Department as well as the Kavli Institute of Nanoscience Delft. J.Y.C. acknowledges NSF-DMR 2209804 and Welch AT-2056-20210327 for partial support of this work.

**Author contributions**

Y.W. and H.W. wrote the majority of the manuscript. M.N.A and J.Y.C are the principle investigators. All authors contributed to the writing of the manuscript.

**Competing interests**

The authors declare no competing interests.

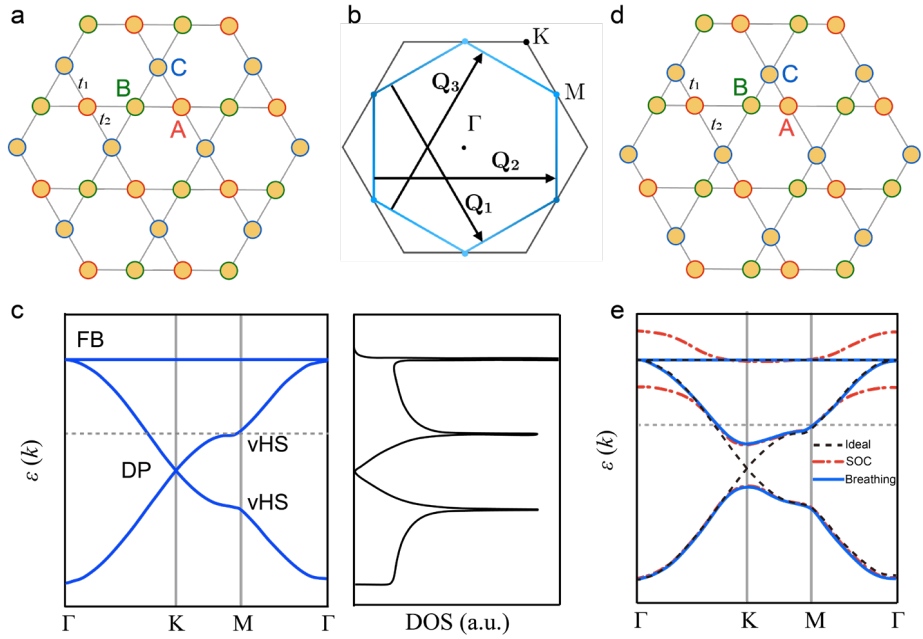

**Figure 2.** Lattice structure and electronic band property of kagome lattice. **a**. Basic structure of ideal kagome lattice. **b**. Schematic of the Fermi surfaces in the Hexagonal Brillouin zone, the Fermi surfaces is nested by three ordering wave factors $Q_i$, *i*=1,2,3. The image is from Ref. [127][127]. **c**. Schematic of the band structure of ideal kagome lattice and corresponded DOS, the van Hove singularity (vHS), Dirac point (DP), and flat band (FB) are labeled. **d.** Structure of breathing kagome lattice, where the hopping amplitude $t_1$ in upward triangles is different with $t_2$ in downward triangles. **e.** Schematic band structures for ideal kagome lattice(black dashed line), influence of SOC (red dashed line) and breathing kagome lattices(blue solid line).The image is based on the result in Ref.[51][51].

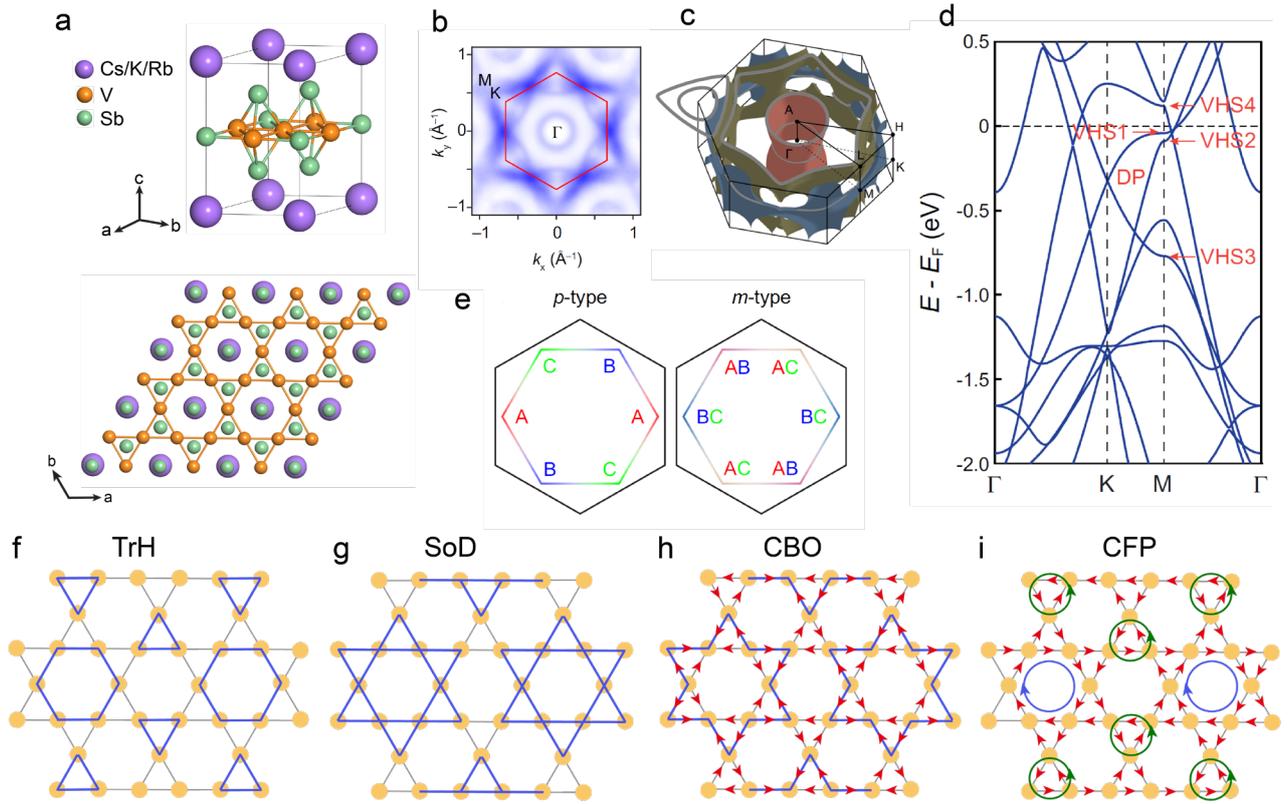

**Figure 3.** Lattice structure and electronic band property of AV$_3$Sb$_5$(A=Cs, K, Rb). **a**. Basic structure of AV$_3$Sb$_5$ **b**. ARPES measured Fermi surface of KV$_3$Sb$_5$. **c**. Fermi surfaces for undistorted CsV$_3$Sb$_5$ in the Hexagonal Brillouin zone. The image is from Ref. [87][87]. **d**. Band structure of CsV$_3$Sb$_5$, the image is obtained based on Ref [63][63]. **e**. Two kinds of vHSs with distinct sublattices decoration, the p-type vHS is sublattice pure, m-type vHS has mixed sublattice feature. **f-i** are possible charge order state predicted for AV$_3$Sb$_5$ family. **f** and **g** are charge orders with tri-hexagonal structure, and Star of David structure respectively; **h** is one kind of predicted charge bond order breaking time-reversal symmetry; **i** is one kind of chiral flux phase breaking TRS.

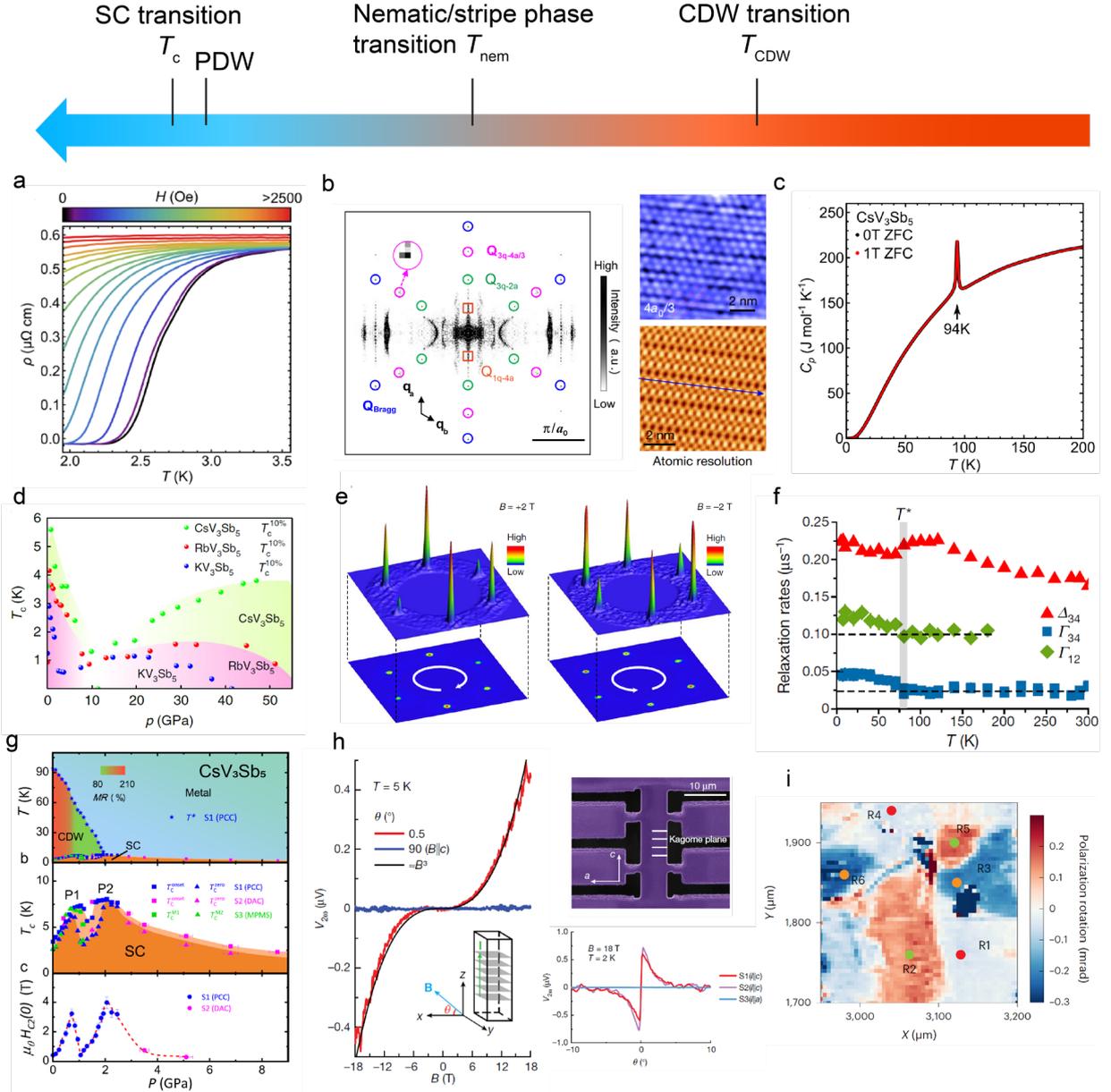

**Figure 4.** Representative physical properties observed in AV$_3$Sb$_5$. **a**. Superconductivity at low temperature. The image is from Ref.[30][30] **b**. $2a_0\times 2a_0$ CDW order ($Q_{3q-2a}$=1/2$Q_{Bragg}$), $4a_0$ CDW order ($Q_{1q-4a}$=1/4$Q_{Bragg}$), and PDW order ($Q_{3q-4/3a}$=3/4$Q_{Bragg}$) measured by STM/STS in CsV$_3$Sb$_5$. The images are from Ref.[29][29]. **c**. CDW transition measured in CsV$_3$Sb$_5$, the image is from Ref.[30][30]. **d and g** are double-dome superconducting phase diagram under high pressure and relatively low pressure region, respectively. The image is from Ref.[130,135][130,135] **e**. Chiral charge order detected by STM in KV$_3$Sb$_5$. The image is from Ref.[69][69] **f**. Signal of time-reversal symmetry broken measured by high resolution muon spin scattering in KV$_3$Sb$_5$. The images are from Ref.[119][119]. **h**. Chiral transport with second harmonic generation (SHG) signal detected in CsV$_3$Sb$_5$. The device is shown in the top right. The right bottom panel is angle dependence of SHG signal, which is switchable by the out-of-plane magnetic field. The image is from Ref.[128][128] **i**. Three nematic domains detected in RbV$_3$Sb$_5$ by birefringence measurement. The red, green and blue regions show the domains. The image is from Ref.[78][78]

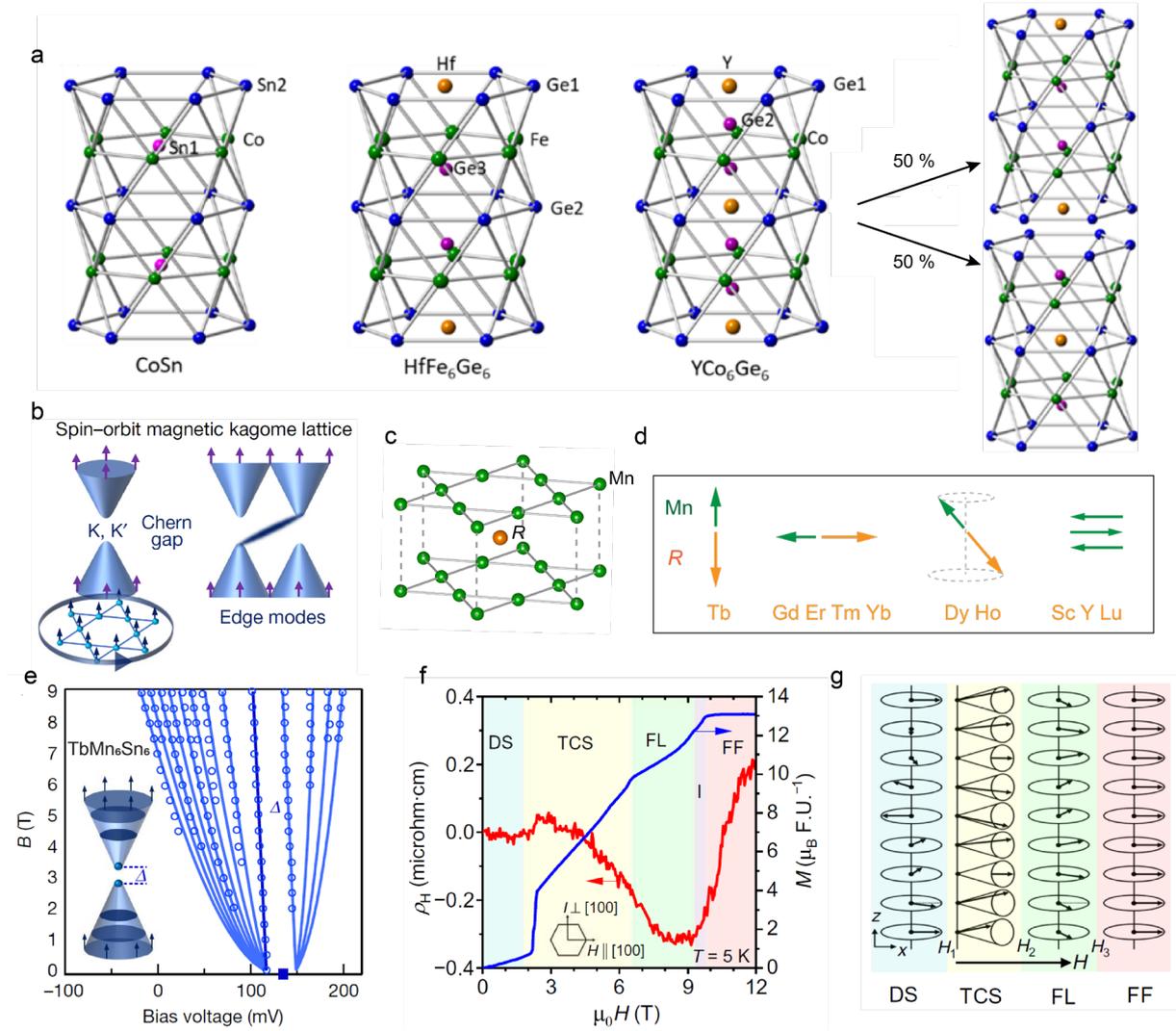

**Figure 5.** Crystal structures and physical properties in "166" compounds. **a**. Crystal structures of CoSn, HfFe$_6$Sn$_6$, YCo$_6$Ge$_6$ types. The HfFe$_6$Sn$_6$ structure type is an "ordered" variant of a stuffed CoSn structure type. The YCo$_6$Ge$_6$ structure type is a "disordered" variant of a stuffed CoSn structure type. For clarity, the disorder of the YCo$_6$Ge$_6$ structure type is broken down into two possible stuffing arrangements. The image of **a** is from Ref.[41][41]. **b**. The schematic band structure of spin-orbit-coupled magnetic kagome lattice, where the spin-polarized Dirac bands opens a Chern gap with edge mode arising in the gap. **c**. Two kagome lattices in a unit cell of $R$Mn$_6$Sn$_6$. **d**. The magnetic state of $R$ atom (orange arrow) and Mn (blue arrow) in different $R$Mn$_6$Sn$_6$ compounds. **e**. STM measured Landau fan data (open circles) in TbMn$_6$Sn$_6$, which is fitted with the spin-polarized and Chern gapped Dirac dispersion (solid line). The inset shows the schematic of Landau quantization of Chern-gapped Dirac band. **f**. Hall resistivity and magnetization measured in YMn$_6$Sn$_6$. The magnetic structures changes from distorted spiral (DS) state, to transverse conical spiral (TCS), fan-like (FL) state, and forced ferromagnetic (FF) state when increasing magnetic field. **g**. Schematic of different field-induced magnetic structures.. The images of **b** and **e** are from Ref.[35][35]. The images of **f** and **g** are from Ref.[187][187].

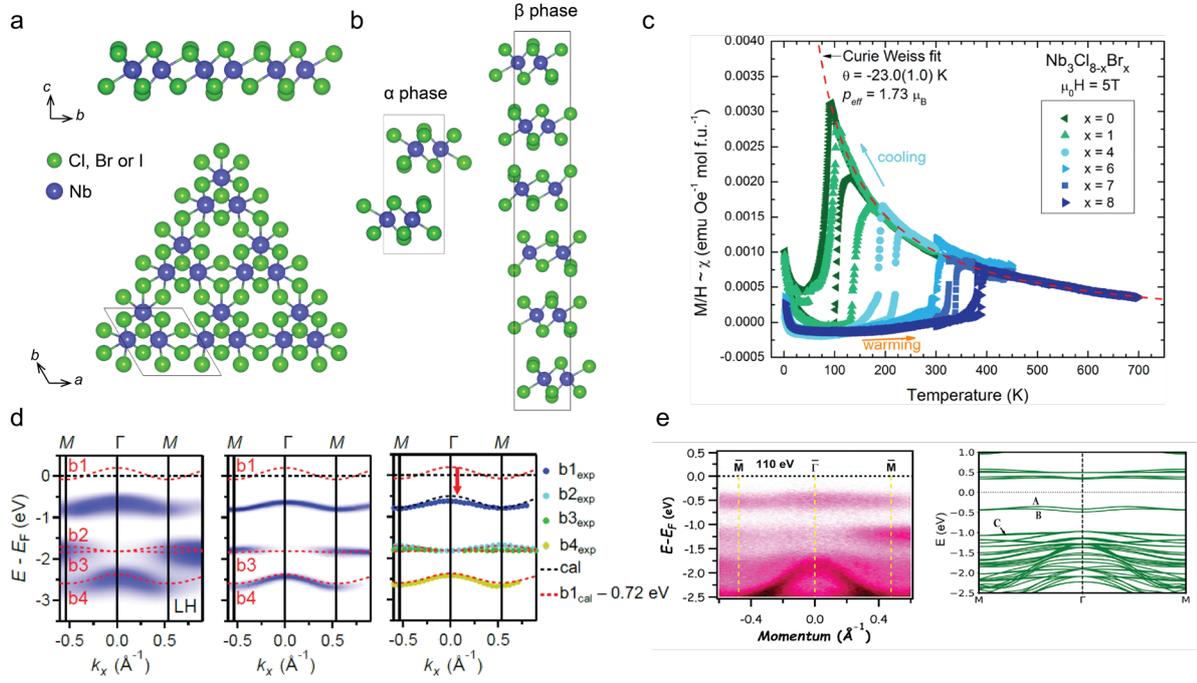

**Figure 6**. Crystal structure and physical properties of $Nb_3X_8$. **a**. Single layer structure of $Nb_3X_8$ and corresponded *ab* plane structure. (b) Crystal structures for α ($P\bar{3}m1$) and β ($R\bar{3}m$) phases of $Nb_3X_8$. (c) Temperature dependent magnetic susceptibility of $Nb_3Cl_{8-x}Br_x$. The image is from Ref.[37][37] (d) Flat band and Mott insulator properties of $Nb_3Cl_8$ bulk measured by ARPES. The left panel is measured with linearly horizontal polarizations of the incident light, the middle panel is corresponded intensity plots of the curvature, the right panel is comparison of experiments with theoretical calculations. The images are from Ref.[40][40] (e) Flat band of $Nb_3I_8$ bulk measured by ARPES, the right panel is corresponded calculated band structure. The images are from Ref. [228][228].

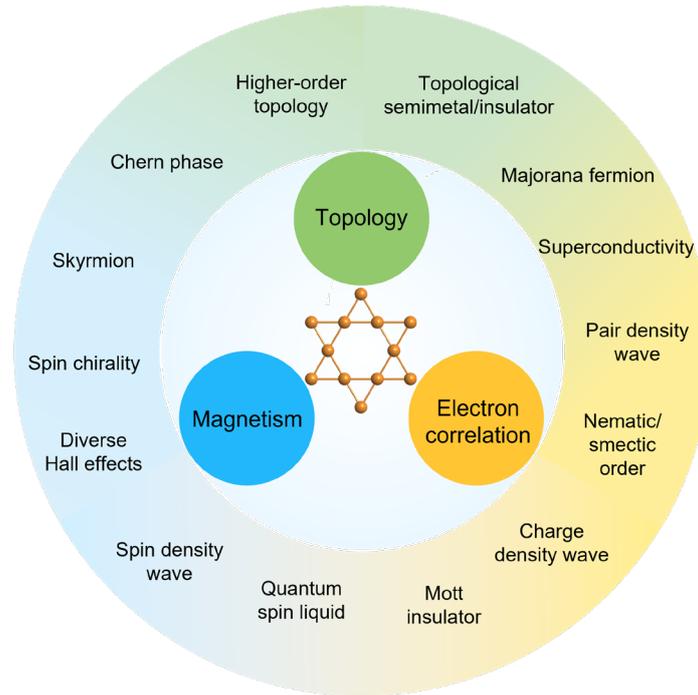

Figure 7. Diverse physical properties in kagome lattice.

Table 1| Properties of Kagome materials

| Materials | Low Temperature Magnetism | Topological | Electron Correlation | Phase transitions |
|---|---|---|---|---|
| $AV_3Sb_5$ (A=Cs, K, Rb) | No long range order but TRSB reported[78,119,120] | Dirac semimetal[32,62] | Weak e-e correlation[84], vHS[62,63], flat band[60] | Structure distortion[86,87], CDW[69], PDW[29], nematic/stripe orders[28,73], SC[30] |
| $RMn_6Sn_6$ | Ferrimagnetic: R=Tb, Dy, Ho, Er (conical magnetic order in R= Dy, Ho)[185] Antiferromagnetic: R= Tm[185] Antiferromagnetic, double-flat-spiral: R=Y, Sc, Lu, Er[187–189] Ferromagnetic: R=Li, Mg, Ca[191,192] | Chern phase: R=Tb[35], Gd-Er[202], proposed for R=Y[204] Dirac cone in $YMn_6Sn_6$[204] | flat band and vHS reported in $YMn_6Sn_6$ | Magnetic transitions |
| $RMn_6Ge_6$ | Ferromagnetic: R=Nd, Sm[179] Antiferromagnetic: R=Dy-Yb, Sc, Y, Lu, Gd[177,193] | Not reported | Not clear | Magnetic transition |
| $RFe_6Sn_6$, $RFe_6Ge_6$ R=Dy-Yb, Sc, Y, Lu | Mostly antiferromagnetic[177,178] | Not reported | Not clear | Magnetic transitions |
| $RV_6Sn_6$ | Non-magnetic: R= Y[194] Weak paramagnetic: R=Sc[195] R=Gd: weak, field dependent magnetism[194] | Dirac cone in R=Gd, Ho[201,202] | vHS in R=Gd, Ho[201,202] | Structural and CDW transitions in $ScV_6Sn_6$[195] Magnetic transition in $GdV_6Sn_6$[194] |
| $MoCo_6Ge_6$ $Yb_{0.5}Co_3Ge_3$ | Paramagnet[196] Weak magnetism, possible magnetic transition around 20 K[41,42] | Not reported Not reported | Not clear Not clear | Structural transition[196] Structural transition[42] |
| $Nb_3X_8$ (X=Cl, Br, I) | X=Cl, Br, paramagnetic to nonmagnetic transition (bulk crystal)[37] Thin film: not clear | Not reported | Strong correlation Flat band[40,223,228] Mott insulator[39,40] | Structural and magnetic transition |
| FeGe FeSn CoSn | Antiferromagnetic[230] Antiferromagnetic[15] Paramagnetic[16] | Not reported Dirac cone[15] Dirac cone[16] | vHS[230] Flat band[15] Flat band[16,17] | FeGe[230]: Structural, CDW and magnetic transitions |
| $Mn_3Ge$, $Mn_3Sn$ | Chiral antiferromagnetic[20,22] | Weyl semimetal[21] | Not clear | Magnetic transition |
| $Co_3Sn_2S_2$ | Ferromagnetic[14] | Weyl semimetal[13,14] | Flat band[235] | Magnetic transition |
| $Fe_3Sn_2$ | Noncollinear ferromagnetic, skyrmion[18,19] | Massive Dirac point[236] | Flat band[237] | Magnetic transition |
| $RT_3X_2$ | Mostly weak magnetism[213–215,233,238,239] | Dirac cone proposed[214,234] | Flat band, vHS proposed[214,234] | SC in many compounds, e.g. $LaRu_3Si_2$[213,240], $LaIr_3Ga_2$[214], $YRu_3Si_2$[215], $LaRh_3B_2$[239], $CeRh_3B_2$[239] |

| Terms | Abbreviation |
|---|---|
| Superconductivity | SC |
| Charge density waves | CDW |
| Spin density waves | SDW |
| Pair density waves | PDW |
| Spin orbit coupling | SOC |
| Van Hove singularities | vHS |
| Density of states | DOS |
| Angle-resolved photoemission spectroscopy | ARPES |
| Nuclear magnetic resonance | NMR |
| Star of David | SoD |
| Tri-Hexagonal | TrH |
| Muon spin spectroscopy | μSR |
| Time-reversal symmetry breaking | TRSB |
| Charge bond order | CBO |
| Nuclear quadrupole resonance | NQR |
| Transverse conical spiral phase | TCS |